\documentclass[twocolumn,showpacs,preprintnumbers,amsmath,amssymb]{revtex4}

\begin{document}
\title{Secure quantum string seal exists}
\author{Guang Ping He}
\affiliation{Department of Physics and Advanced Research Center,
Zhongshan University, Guangzhou 510275, China}

\begin{abstract}
It was claimed that all quantum string seals are insecure [H. F.
Chau, quant-ph/0602099]. However, here it will be shown that for
imperfect quantum string seals, the information obtained by the
measurement proposed in that reference is trivial. Therefore
imperfect quantum string seals can be unconditionally secure.
\end{abstract}

\maketitle

\newpage

The idea of quantum seal was first proposed by
Bechmann-Pasquinucci \cite {sealing}, who also found that secure
perfect quantum seal is impossible \cite{impossibility}. Soon it
was found by He \cite{He} that security bounds also exist for
imperfect quantum bit seal. He proposed an explicit cheating
strategy with which the reader can decode the sealed message with
exactly the same reliability as that of the honest measurement,
while the cheating can escape from being detected with a
probability which cannot be made arbitrarily small. Shortly after
that, Chau studied a less general model of
imperfect quantum bit seal (which is a special case of He's model where $%
\{g\notin G_{0}\cup G_{1}\}=\emptyset $ and $\alpha $ is known to
Bob), and proposed a measurement strategy which will cause less
detectable disturbance to the seal if the reader sacrifices the
reliability of the message he decoded \cite{Chau}.

At the same time, an imperfect quantum string seal scheme was
proposed \cite {String}. Though the security of each single bit in
the sealed string is still restricted by the security bounds of
imperfect quantum bit seal \cite {He,Chau}, the whole string can
be secure in the sense that decoding much of the bits will result
in a non-trivial probability of being detected. But recently, it
was claimed that all quantum string seals are insecure \cite
{Insecure}. The claim was based on a measurement strategy which is
said to be able to obtain non-trivial information on the sealed
message, and escape verifier's detection at least half of the
time. However, it will be proven below that for imperfect quantum
string seals, the information obtained by this measurement is
trivial. Therefore the security of imperfect quantum string seal
\cite{String} can not be denied.

According to Ref. \cite{Insecure}, the measurement strategy it
proposed can enable Bob to distinguish the sealed message with the
probability $p<p_{\max }$. Here $p_{\max }$\ is a parameter
determined by the quantum string seal scheme chosen by Alice. But
we must notice an important difference between imperfect quantum
bit seal and string seal. That is, $p_{\max }$\ needs not to have
a large value in the latter. The readability requirement of
imperfect quantum string seal is merely to enable Bob to obtain a string $%
b^{\prime }$, which matches the original $n$-bit string $b$ sealed
by Alice except with a small error rate $\varepsilon $ (i.e. the
Hamming distance between $b$ and $b^{\prime }$\ is not greater
than $\varepsilon n$). In other words, each bit of $b^{\prime }$
can be different from the corresponding bit of $b$ with the
probability $\varepsilon $. Note that this requirement never
implies that the probability for $b^{\prime }$ to be
exactly equal to $b$ (i.e. the probability $p_{\max }$) needs to be $%
(1-\varepsilon )$. In fact, even for a very small $\varepsilon $,
the number
of possible $b^{\prime }$ can be numerous. Meanwhile, the probability $%
p_{\max }$ for $b^{\prime }=b$ to occur is extremely small. Taking
the
scheme proposed in Ref. \cite{String} for example, the error rate is $%
\varepsilon \equiv \sin ^{2}(\Theta /n^{\alpha })$, where $\Theta $\ ($%
0<\Theta \ll \pi /4$) and$\ \alpha $\ ($0<\alpha <1/2$) are fixed
constants.
We can see that the readability of the sealed string can be very high since $%
(1-\varepsilon )$ can be made arbitrarily close to $1$ by
increasing $n$. On the other hand, the order of magnitude of
$p_{\max }$ is $(1-\varepsilon )^{n}\backsim (1-\Theta
^{2}/n^{2\alpha })^{n}$, which can be made arbitrarily small as
$n$ increases.

So we can see that with the measurement strategy proposed in Ref.
\cite {Insecure}, the probability $p$ for Bob to distinguish
faultlessly the original sealed string $b$ is very small.
Therefore the most important parameter characterizing the amount
of information obtained by the measurement strategy is not the
probability $p$. Instead, it is the reliability of Bob's output
string $b^{\prime }$ in the rest $(1-p) $\ cases, i.e., how much
information of $b$ is reflected by $b^{\prime }$. Here we give a
brief evaluation with a simple string seal scheme. Suppose that
the sealed state for the message $i^{\prime }$ is
\begin{equation}
\left| \tilde{\psi}_{i^{\prime }}\right\rangle
=\sum\limits_{j^{\prime }}\lambda _{i^{\prime }j^{\prime }}\left|
\psi _{j^{\prime }}\right\rangle _{B}.
\end{equation}
The expression of the proposed measurement in Ref. \cite{Insecure}
(see Eq. (18) of that reference) can be rewritten as
\begin{equation}
M_{i}=(a+b)\left| i\right\rangle \left\langle i\right|
+a\sum_{j\neq i}\left| j\right\rangle \left\langle j\right| .
\end{equation}
Applying this measurement on the sealed state, we have
\begin{equation}
M_{i}\left| \tilde{\psi}_{i^{\prime }}\right\rangle =(a+b)\lambda
_{i^{\prime }i}\left| i\right\rangle +a\sum_{j\neq i}\lambda
_{i^{\prime }j}\left| j\right\rangle .
\end{equation}
Thus the probability for the message $i^{\prime }$ to be decoded
as $i$ is
\begin{eqnarray}
p_{i^{\prime }i} &=&a^{2}+(2ab+b^{2})\lambda _{i^{\prime }i}^{2}
\nonumber
\\
&=&\frac{1-\nu }{N}+\nu \lambda _{i^{\prime }i}^{2}.
\end{eqnarray}
Taking $\nu =1/2$, as suggested by Ref. \cite{Insecure}, it gives
\begin{equation}
p_{i^{\prime }i}=\frac{1}{2N}+\frac{\lambda _{i^{\prime
}i}^{2}}{2}.
\end{equation}
This equation means that any one of the $N$ possible messages has
at least the probability $1/(2N)$ to be decoded as message $i$,
even if its content has nothing to do with $i$ at all. In other
words, whenever Bob obtains a message via the measurement
strategy, there is a probability $N\cdot 1/(2N)=1/2$ that the
original message can be anything, i.e., the amount of information
he obtained is zero. Hence we can see that what was obtained in
Ref. \cite{Insecure} is simply a natural the result -- Bob obtains
useful information only with less than $1/2$ chance, therefore the
disturbance on the sealed state is also less than half. This is
consistent with the security proof of the quantum string seal
\cite{String}, that reading less bits of the string will stand
less chances to be detected. It does not mean that unconditionally
secure imperfect quantum string seal is impossible. Instead, it
provides a clear example of how the quantum string seal remains
secure against coherent measurements.

At last, I would like to provide a simple measurement strategy
which is as efficient as the one proposed in Ref. \cite{Insecure},
while no quantum computer is needed. Bob can simply toss a coin to
decide his action. At half of the cases he performs the honest
measurement (which generally requires individual measurement only)
and reads the string, while at the other half of the cases he does
nothing. Clearly he can obtain the sealed message with the
probability $1/2$, while at the other half of the time the sealed
state remains perfectly intact so he will not be detected.

This work was supported in part by the Foundation of Zhongshan
University Advanced Research Center.

\end{document}